\documentstyle[twocolumn,prb,aps]{revtex}
\begin{document}
\draft
\title
{Restrictions on the coherence of the ultrafast optical emission from an
electron-hole pairs condensate}
\author{A.Olaya-Castro$^{1}$, F.J.Rodr\'{\i}guez$^{1}$, L.Quiroga$^{1}$
and C.Tejedor$^{2}$}
\vspace{.05in}
%
\address
{$^{1}$ Departamento de F\'{\i}sica, Universidad de Los Andes, A.A.
4976,
Bogot\'a D.C., Colombia\\
$^{2}$ Departamento de F\'{\i}sica Te\'orica de la Materia Condensada,
Universidad Aut\'onoma de Madrid,
Madrid, Spain}
%
\maketitle

\begin{abstract}
We report on the transfer of coherence from
a quantum-well electron-hole condensate to the light it emits. 
As a function of density, the coherence of the electron-hole pair system 
evolves from being full for the low density Bose-Einstein condensate to a 
chaotic behavior for a high density BCS-like state. This degree of coherence is 
transfered to the light emitted in a damped oscillatory way in the ultrafast regime.
Additionally, the photon field exhibits squeezing properties 
during the transfer time.
Our results suggest new type of ultrafast experiments
for detecting electron-hole pair condensation.
\end{abstract}
\vspace{.25in}


\narrowtext

The generation of quantum coherence and 
entanglement is of much interest
for testing certain aspects of the nonlocal predictions of
quantum mechanics as well as for applications in the emerging field
of quantum information
processing. The recent achievement of Bose-Einstein condensation 
(BEC) in dilute atomic systems\cite{anderson}
has triggered a great interest 
in looking for such quantum correlations
with massive particles.\cite{grifith} 
Furthermore, the manipulation of the interaction of light with massive particles is crucial
for controlling processes which transfer coherence and/or
entanglement between radiation and matter. A condensed 
matter system being a candidate for this goal is
an electron-hole pair condensate.\cite{keldysh,blatt}
Several groups have directed their experimental efforts to produce this
collective state in semiconductors.\cite{wolfe,mendez,butov}
Recently, long lifetime indirect excitons, in both
real and ${\bf k}$ spaces, have been proposed as the most 
robust entities towards condensation.\cite{lozovik1,lozovik2,butov3}
Hence, it is natural to inquire which signatures may be expected
to be transfered from the electron-hole pair condensate to emitted photons.
The transfer of coherence is an ultrafast process that can be analyzed 
by means of coherent control techniques as those recently applied to 
semiconductor nanostructures. This work is a first step in this direction.

Coherence of a quantum system is associated with the observation of
interference effects which can be described by first- and higher-order correlation
functions as stated by Glauber\cite{glauber}. 
Most of the studies of the light emitted from an electron-hole condensate that 
have been proposed to date rely on the lowest-order fluctuations in photon 
counting experiments, {\it i.e.} intensity measurements instead of 
the correlation functions.\cite{sham,inagaki}
It has been previously shown that at low density the emitted light
should be in a coherent state.\cite{joaquin} However, we shall demonstrate here
that this perfect coherence transfer is only possible asymptotically.
The proper way to quantify both the amount and dynamics of transfered 
coherence to the radiation field is by considering the correlations 
between photons as a function of time.\cite{laikhtman}
Recent proposals\cite{imamoglu} may bring this kind of photon
counting experiments within reach. 
The main aim of this work
is to explore time-dependent higher-order coherence properties of photons emitted
from a collective electron-hole pair state in an ultrafast time scale.
Our results show that full coherence transfer takes
a time 
ranging from hundred of femtoseconds to a few picoseconds.
During this transfer
time squeezing of photons may be achieved.

Since we are interested
in the quantum effects we restrict ourselves to study ground-state
properties (zero temperature) and let the pair density change.
We consider indirect semiconductor quantum wells 
where electrons and holes
are spatially separated by an interlayer distance $d$.
The confinement in the $z$ direction is sufficiently strong so that
we ignore excitations in that direction.
The Hamiltonian of the system includes the kinetic energy and all the 
interactions among electrons and holes,
the free (quantized) electromagnetic field and the interaction
between the radiation and the electron-hole system.
The system's initial state is assumed to be a product of an empty 
radiation field state and a condensate electron-hole state, 
$|0_{photons}\rangle \times \prod_{\vec k} {(u_{\vec k}+v_{\vec k}e_{\vec
k}^{\dag}h_{-{\vec k}}^{\dag}})|0\rangle$
with ${\vec k}$ being a two-dimensional wavenumber vector,
$u_{\vec k}$ and $v_{\vec k}$ satisfy the normalization
condition  $u_{\vec k}^2+v_{\vec k}^2=1$.
$e_{\vec k}^{\dag}$ ($h_{\vec k}^{\dag}$)
is the electron (hole) creation operator and $|0\rangle$ denotes the
semiconductor ground state.
The variational BCS-like function has enjoyed considerable success in the
description of stationary properties of electron-hole systems.\cite{nozieres}
It captures the essential electron-hole pairing correlations in the low
as well as high pair density, $n$, limits,
although it does not
describe possible collective modes in the condensate.
We also assume that in the ultrafast coherence transfer period
just a few bunch of photons is emitted, so that $n$ doesn't change significantly
and thus we take it as a constant.
The time-dependent optical coherence is obtained via photon
operators determined by pairs evolving under the action of
the electron-hole Hamiltonian in the free
Bogoliubov quasiparticles approximation\cite{comte} (quantum fluctuation
terms are neglected). These basic assumptions are indeed satisfied
for transfer times below a few picoseconds 
and more interestingly they also yield to the 
correct stationary limit.
Coefficients $u_{\vec k}$ and $v_{\vec k}$
are found from the self-consistent solution of the BCS gap
equation for given chemical potential
$\mu$ and $n$.
In what follows, energy is measured in 2D Rydbergs 
${\mathcal{R}} _0$, length in 2D Bohr radius $a_0$, and time in units of 
$\tau_0={\mathcal{R}} _0^{-1}$. 

Since an electron-hole condensate corresponds to a coherent macroscopic 
polarization of the electronic states, 
we start by considering the polarization correlation functions (PCFs).
The first- and second-order PCFs are defined
as $G_P^{(1)}(t_1;t_2)=\langle P_{\vec 0}^\dagger(t_1)P_{\vec 0}(t_2)\rangle$  
and $G_P^{(2)}(t_1;t_2)= \langle P_{\vec 0}^\dagger(t_1)
P_{\vec    0}^\dagger(t_2)P_{\vec 0}(t_2)P_{\vec 0}(t_1)\rangle$ respectively, 
where   $P_{\vec  q}=\sum_{\vec  k}h_{{\vec  q}-{\vec
k}}e_{\vec k}$ is the polarization field. Due to the long wavelength
of emitted photons, only ${\vec q}={\vec 0}$
modes are important. PCFs in the BCS state are better expressed in terms of
the so-called normal and anomalous (or pair) Green's functions
\cite{laikhtman}, defined by ${\mathcal{G}}(t_1-t_2)=\sum_{\vec k}\langle
e_{\vec k}^\dagger(t_1)e_{\vec k}(t_2)\rangle\langle
h_{-{\vec k}}^\dagger(t_1)h_{-{\vec k}}(t_2)\rangle$
and ${\mathcal{F}}(t_1-t_2)=\sum_{\vec k}\langle h_{-{\vec k}}(t_1)
e_{\vec k}(t_2)\rangle$, respectively. The first-order PCF can be written as
$G_P^{(1)}(t_1-t_2)={\mathcal{G}}(t_1-t_2)+|{\mathcal{F}}(0)|^2$.
It should be noted that the polarization field remains always in a
stationary state in such a way that its correlation and coherence
functions are only dependent on $\tau=t_2-t_1$ but not on $T=(t_1+t_2)/2$.
The first-order polarization coherence function is expressed as
\begin{eqnarray}
g_P^{(1)}(\tau)=
\frac{G_P^{(1)}(\tau)}{G_P^{(1)}(0)}=\frac{\sum_{\vec k}|v_{\vec
k}|^4e^{iE_{\vec
k}\tau}+[\sum_{\vec k}
u_{\vec k}v_{\vec k}]^2}{\sum_{\vec k}|v_{\vec k}|^4+[\sum_{\vec k}
u_{\vec k}v_{\vec k}]^2}
\end{eqnarray}
where $E_{\vec k}$ is the excitation energy of Bogoliubov quasiparticles.  
Similarly, the second-order polarization coherence function is
\begin{eqnarray}
g_P^{(2)}(\tau)=1+
|g_P^{(1)}(\tau)|^2-\frac{|{\mathcal{F}}(0)|^4}{|G_P^{(1)}(0)|^2}
\end{eqnarray}
where $|{\mathcal{F}}(0)|^2=\left[\sum_{\vec k} u_{\vec k}v_{\vec
k}\right]^2$. It must be stressed that 
the second-order coherence function for the polarization, $g_P^{(2)}(0)$, 
equals 1 for a coherent state whereas it is 2 for a chaotic field.\cite{loudon}

Our results for the second-order polarization field coherence, for $d=0$ and 
different densities are shown in Fig. 1. Clearly, a full coherent behavior, 
in the Glauber sense, appears in the low density limit. In this case 
$v_k^2<<1$, yielding to $|g_P^{(1)}(\tau)|=|g_P^{(2)}(\tau)|\simeq 1$, as 
expected for an ideal electron-hole BEC. By contrast, as the density increases 
$g_P^{(2)}(0)$ becomes greater than 1, going up to a maximum value of 2.  
As a function of $\tau$, $g_P^{(2)}$ 
decreases from its initial value in an        
oscillatory way, with a period determined by $E_{{\vec k}=0}$.
At high densities $g_P^{(1)}(\tau)$ decays to zero and
$g_P^{(2)}(\tau)$ decays to one in a faster oscillating way with a frequency 
approaching that of the system's Fermi energy. In the very high density limit, 
where $v_{\vec k}=1$ for $k$ smaller than the Fermi wave vector ${\vec k_F}$ 
and 0 otherwise, the anomalous term vanishes yielding to
$g_P^{(1)}(\tau)=[e^{-i\epsilon _F\tau}sin(\epsilon _F\tau)]/\epsilon _F\tau$ and
$g_P^{(2)}(\tau)=1+|g_P^{(1)}(\tau)|^2$ with $\epsilon _F=k_F^2/2$. 
Clearly, $g_P^{(2)}(0)=2$, corresponding to a non-interacting fermion field, 
and it decays, as a function of $\tau$, in an overdamped way. 
 
Next we discuss the coherence properties of the emitted light in order to 
study the transfer of coherence from the condensate. First- and 
second-order correlation functions are
$G_{\vec q}^{(1)}(t_1;t_2)=\langle C_{\vec q}^\dagger(t_1)C_{\vec
q}(t_2)\rangle$ and $G_{\vec q}^{(2)}(t_1;t_2)=
\langle C_{\vec q}^\dagger(t_1)C_{\vec q}^\dagger(t_2)C_{\vec q}(t_2)C_{\vec
q}(t_1)\rangle$ where $C_{\vec q}^\dagger$ ($C_{\vec q}$) is the photon
creation (annihilation) operator. For ${\vec q=0}$ photons,
$G^{(1)}(t_1;t_2)=|M_o|^2e^{-i\omega_0(t_2-t_1)}\int_{0}^{t_1}dt_a\int_{0}
^{t_2}dt_b\, f(t_b-t_a)$ with $f(t_b-t_a)=e^{i(\omega_0-E_g-\mu)(t_b-t_a)}
G_P^{(1)}(t_b-t_a)$, $\omega_0$ the frequency at which the emitted light is 
filtered, $E_g$ the energy gap and $M_0$ the coupling 
between the photons and the electron-hole pairs. Similarly, 
$G_{\vec q}^{(2)}(t_1;t_2)$ is closely related to
$G_P^{(2)}(t_1;t_2)$. From these relations between
polarization and radiation correlation functions the
coherence transfer from the former to the latter can be obtained.
The first-order correlation function can be written as
$G^{(1)}(T,\tau)= N(T,\tau)+ A(T,\tau)$ where the normal and the
anomalous dimensionless contributions are given by
\begin{eqnarray}
N(T,\tau)=2M_0^2e^{-i\omega_0\tau}\sum_{\vec k}
|v_{\vec k}|^4e^{-i(\omega+E_k)\tau}\\ \nonumber
\times \left[\frac{\cos [(\omega+E_k)\tau/2]
-\cos[(\omega+E_k)T]}{(\omega+E_k)^2}\right]
\end{eqnarray}
and
\begin{eqnarray}
A(T,\tau)=2M_0^2|{\mathcal{F}}(0)|^2 e^{-i(\omega+2\omega_o)\tau/2}
\\ \nonumber \times \left[\frac{cos(\omega \tau/2)-cos(\omega T)}{\omega ^2}\right]
\end{eqnarray}
respectively, with $\omega=\omega_0-E_g-\mu$. 
The first- and second-order coherence functions are then
\begin{eqnarray}
g^{(1)}(T,\tau)=\frac{G^{(1)}(T,\tau)}
{\sqrt{G^{(1)}(t_1;t_1)G^{(1)}(t_2;t_2)}}
\\\nonumber
\end{eqnarray}
and
\begin{eqnarray}
g^{(2)}(T,\tau)=1+|g^{(1)}(T,\tau)|^2-
\frac{|A(T,\tau)|^2}{G^{(1)}(t_1;t_1)G^{(1)}(t_2;t_2)}
\end{eqnarray}
respectively. As for the polarization, the anomalous contribution
determines whether or not the light is second-order coherent. 
The steady-state expressions can be quickly obtained by
letting $T>>\tau$ which leads to $G^{(1)}(t_1;t_1)\simeq
G^{(1)}(t_2;t_2)\simeq G^{(1)}(T,\tau=0)$.

First- and second-order coherence do
not depend on the efficiency of the detector used in photon
counting experiments nor on
the coupling $M_0$. Moreover, the amount of coherence depends 
only on the system's density. Turning this around, we can use the coherence
information to determine the electron-hole density.

One expects the emitted light to be fully coherent in the
low-density limit whereas it should be chaotic in the high-density
limit as emitted by an uncorrelated source.
We indeed show that in both cases the asymptotic exact solutions are
recovered. Obviously,
from Eq.(5) it can be drawn that $g^{(1)}(T,0)=1$ for any frequency
and/or electron-hole pair density, 
in agreement
with the standard result according to which any single mode radiation
field is first-order coherent\cite{loudon}. 
Therefore, we restrict ourselves to
second-order coherence properties.  
In the low density limit,
$v_k$ is essentially the hydrogenic ground state wavefunction,
$v_k\simeq \sqrt n\, \phi_{1s}(k)= [2\sqrt{2\pi n}a_0]/[1+k^2a_0^2]^{3/2}$,
$\mu\simeq-{\mathcal{R}} _0$ and the normal contribution vanishes. 
The second-order coherence
function becomes $|g^{(2)}(T,\tau)|\simeq1$, indicating the fact that in a
steady-state situation the light emitted from an electron-hole pair condensate
(BEC state) is second-order coherent, in full agreement with previous
results for excitons.\cite{joaquin} In the very high density limit, where
$\mu\simeq{k_F}^2$,  the anomalous part goes to zero.
The second-order coherence is now
given by $g^{(2)}(T,\tau)=1+|g^{(1)}(T,\tau)|^2$,so that
$g^{(2)}(T,0)\simeq2$ as it corresponds to chaotic radiation.\cite{tesis}

Now we turn to the more general case of arbitrary densities. 
Figure 2 depicts
$g^{(2)}(T,0)$ at $\omega_0=E_g+\mu$, for different densities.
Clearly, two very different behaviors are observed depending on the
time-scale considered. In the short time regime, the emitted
radiation is {\it  partially coherent} since  $g^{(2)}>1$;
this behavior is reinforced as the density increases.
For $T\rightarrow 0$, $g^{(2)}(T,0)$ approaches $g_P^{(2)}(0)$, which
depends only on the system's density and characterizes the 
fluctuations of the macroscopic collective polarization state. In this way
emitted photons could bring well differentiated information on the
system's ground state.

Second-order coherence describes also the tendency  of photons to
arrive in pairs ($g^{(2)}(T,0)>1$) or  rather to be spaced out  in
time ($g^{(2)}(T,0)<1$).\cite{loudon} Our results show
that at a short time scale the first condition is satisfied, 
producing photon bunching. In
a long time scale, the photon bunching effect disappears and
the radiation field becomes {\it  asymptotically coherent}, {\it
i.e.}  $g^{(2)}(T,0)\rightarrow 1$, for any  finite density. This is
due to the fact that the normal contribution, $N(T,0)$, saturates to a
constant value while the anomalous contribution, $A(T,0)$, grows as
$T^2$.\cite{tesis}  It must be stressed that the light
emitted  is coherent  even though the polarization field is
incoherent in the Glauber sense. The time to reach the steady-state
value is longer as the density increases. For systems of interest:
(i) GaAs, ${\mathcal{R}} _0 \simeq 16 meV$, $a_0\simeq 62.5$\AA, and
for a density $3\times10^{10}cm^{-2}$, the stationary regime is
reached roughly after 1 picosecond; (ii) CdS, ${\mathcal{R}} _0\simeq
120 meV$,  $a_0\simeq 12.75$\AA, and for a density
$7\times10^{11}cm^{-2}$  a steady-state situation is reached for a
time on the order of $100$ femtoseconds. In
both cases, $na_0^2=1.3\times 10^{-2}$.

These  results  show  how the coherence of the photon field  at
$\omega_0-E_g=\mu$, evolves from  a partially coherent   behavior,
dominated by the fermionic character of the system, towards a full coherent
behavior, reflecting the system's macroscopic quantum properties.
By contrast, light observed  at  frequencies
such that $\omega_0-E_g<\mu$ evolves as a function of time from a
partially coherent character towards a full chaotic behavior, {\it
i.e.}  $g^{(2)}\rightarrow  2$ (Fig. $2$ inset $(a)$; note that for
plotted densities $\mu<0$). Clearly, a successful coherence transfer
is only possible for the former case but not for the latter one.

Interlayer separation effects are shown in the insets $(b)$ and
$(c)$ of Fig. 2. For increasing $d$ the incoherence of light in the
ultrafast regime becomes more evident and the evolution of
$g^{(2)}(T,0)$ towards its coherent value is slower.  For a fixed
density, $g^{(2)}(T,0)$ when $T\simeq 0$, as a function of $d$,
saturates to a final value of 2, indicating an enhancement of the
chaotic behavior of light.

In order to further characterize the statistical properties of the
emitted radiation, we calculate the variance of the photon field amplitudes
$\hat{X_1}=\frac{1}{2}\{C(T)+C^{\dag}(T)\}$ and
$\hat{X_2}=\frac{1}{2i}\{C(T)-C^{\dag}(T)\}$. 
For radiation emitted by the condensate, these variances are 
$\langle(\Delta\hat{X}_{1,2})^2\rangle =
\frac{1}{4}+\frac{1}{2}N(T,0) \pm \frac{1}{2}Re[\langle \Delta C(T)^2\rangle]$
where, for $\omega_0=E_g+\mu$,
$\langle \Delta C(T)^2\rangle=-2M_o^2 e^{-i2\mu T}
\sum_{\vec{k}}u_k^2v_k^2\frac{\sin^2(E_kT/2)}{E_k^2}$.
In contrast to $g^{(1)}$ and $g^{(2)}$, these variances depend on the 
coupling $M_0$. 
Clearly, squeezed light is possible
only  at moderate low densities, where the normal contribution
is negligible but $u_kv_k$ is still important.
This nonlinear effect is due to interactions between electron-hole pairs, 
in agreement with  results obtained by a simple interacting boson
model\cite{nguyen}. Figure $3$ displays the 
deviation of $\langle(\Delta\hat{X}_1)^2\rangle$ 
from $1/4$ (the coherent state value), for different densities. The amount of 
squeezing, as measured by the most negative value for each curve in Fig. 3,
is a non-monotonically function of the pairs density.
There is a maximum squeezing in one of the 
quadratures every time $\mu T=\pi$.

In summary,  we have shown how coherence transfer, from 
an electron-hole  condensate to the photons it emits,
proceeds as a function of time. The condensate itself presents
different degrees of Glauber coherence depending on its  density and the 
electron-hole layer separation.  A full coherence transfer 
is restricted to light with a frequency given by
$\omega_0=E_g+\mu$ and for times greater than a few hundred femtoseconds.
We also predict light squeezing from a moderate 
low density electron-hole pairs system. These coherence transfer properties should
help  experimentalists  searching  for  evidences of  electron-hole pairs  
condensation in quantum wells. 

The authors acknowledge support from COLCIENCIAS (Colombia) project
No.1204-05-10326, AECI (Spain) and MEC (Spain) contract PB96-0085. 
AOC also acknowledges support from ICFES (Colombia).

\newpage
\centerline{\bf Figure Captions}

\bigskip

\noindent Figure 1: Second-order polarization field coherence as
a function of $\tau$ for 
$d=0$ and
different densities.

\bigskip

\noindent Figure 2: Second-order coherence as a function of $T$ for light emitted
at $\omega_0-E_g=\mu$ and $d=0$ for different densities. Insets: (a) $g^{(2)}(T,0)$
at $\omega_0-E_g=-2.5\mu$; (b) $g^{(2)}(T,0)$ for two different $d$ values and 
(c) $g^{(2)}(0,0)$ for $na_0^2=1.3\times 10^{-2}$ as a function of the 
electron-hole layer separation distance. 

\bigskip

\noindent Figure 3: Time evolution of the photon field amplitude variance 
for radiation emitted at $\omega_0=E_g+\mu$ and $M_0=0.1{\mathcal{R}} _0$.

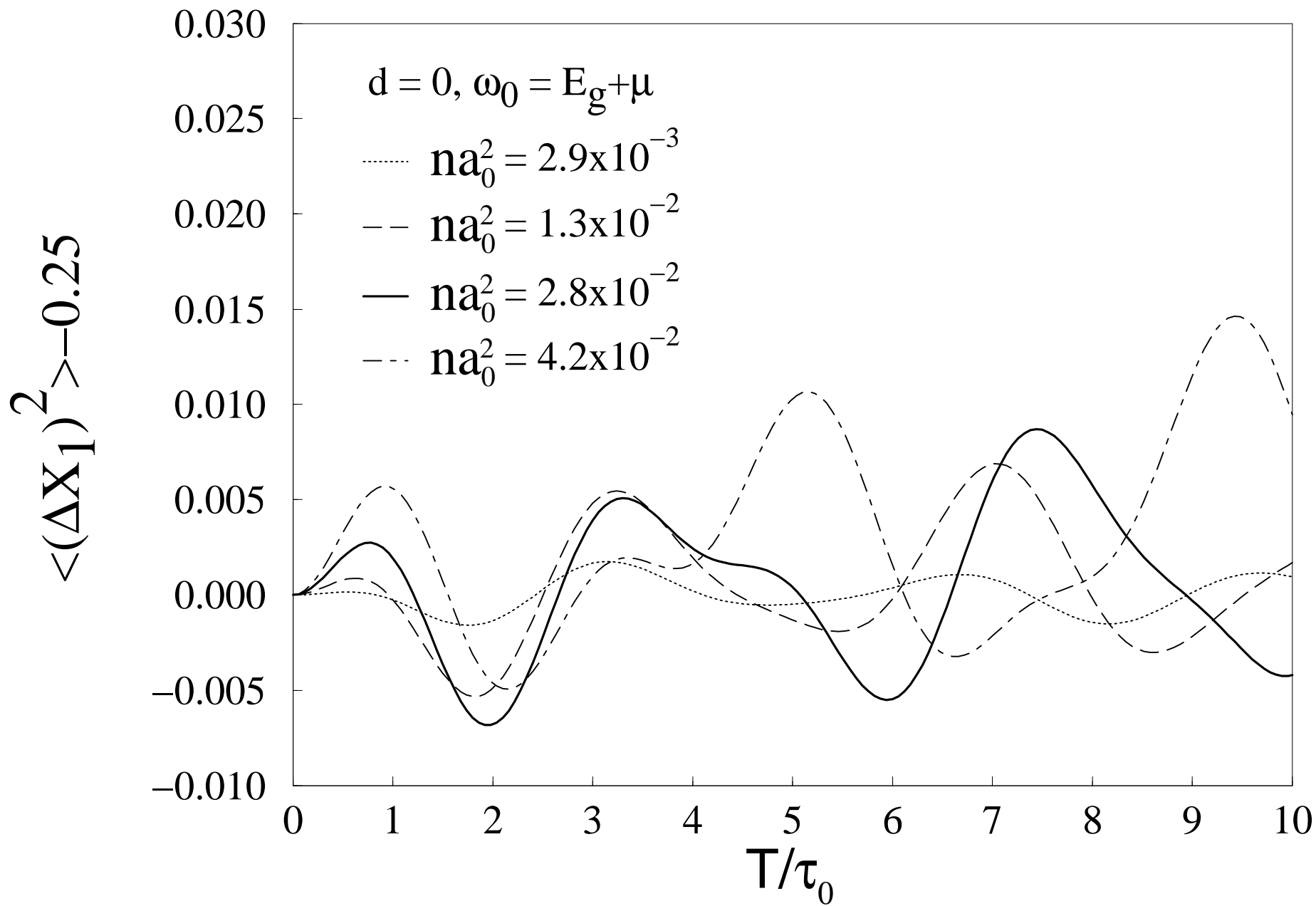
\end{document}